\documentclass[12pt]{iopart}

\usepackage{iopams}  
\usepackage{color}
\usepackage{graphicx}
\begin{document}

\title{Coupling between multiple antennas through a plasma cylinder}
\author{L. Chang$^{1, 2}$, L. P. Zhang$^{1, 2}$, X. G. Yuan$^{1, 2}$, Y. J. Chang$^{1, 2}$, J. H. Zhang$^{1, 2}$, X. Yang$^1$, Y. Wang$^1$, H. S. Zhou$^{1, 2}$, and G. N. Luo$^{1, 2, 3}$}
\address{$^1$Institute of Plasma Physics, HFIPS, Chinese Academy of Sciences, Hefei $230031$, China}
\address{$^2$Science Island Branch of Graduate, University of Science and Technology of China, Hefei $230026$, China}
\address{$^3$School of Nuclear Science and Technology, University of Science and Technology of China, Hefei $230026$, China}

\date{\today}

\begin{abstract}
The coupling physics between multiple antennas separated axially along a plasma cylinder is investigated. Experiments are carried out on a recently built device: Physics ANd Thruster oriented HElicon Research (PANTHER), with an upgrade of second-stage antennas. Mutual induction currents are measured in detail. It is found that the existence of plasma column going through these antennas increase the coupling effects among them significantly. Theoretical analyses from the perspectives of transformer and magnetic permeability and moment confirm the reasonability of this phenomenon. This work is of particular interest for electrodeless plasma source or thruster which employs multiple antennas for ionisation and acceleration. 

\end{abstract}



\maketitle

\section{Introduction}
The scheme of antenna wrapping around a quartz tube has been employed widely for inductively coupling and helicon discharges\cite{Hittorf:1884aa, Boswell:1970aa, Lieberman:2005aa}. This electrodeless feature can lead to long life and high efficiency, due to the absence of electrode erosion and plasma sheath, respectively. To further increase the plasma density and temperature, together with particle energy, multiple antennas were conceived and implemented, such as VASIMR (VAriable Specific Impulse Magnetoplasma Rocket)\cite{Chang-Diaz:2000aa}, HEAT (Helicon Electrodeless Acceleration Thruster)\cite{Shinohara:2014aa} and other experiments\cite{Emsellem:2007aa, Slough:2009aa, Bathgate:2017aa, Yuan:2020aa}. However, the mutual coupling between these coaxial antennas, which are separated axially with a certain distance, through the same plasma cylinder was rarely studied. This coupling can result in significant magnitude of voltage and current on neighbouring antennas and, according to the Faraday's law and the conservation of magnetic flux, they are generally in the opposite direction to those originally provided by their power supplies. With regard to plasma generation, this surely weakens the ionisation rate and heating efficiency. Thereby, it is of great importance to investigate the coupling physics between these antennas, and guide experiments to either reduce the coupling effect or turn the detrimental coupling into beneficial usage. Here, we explore the coupling physics on a recently built device which employs three antennas around a helicon plasma cylinder, and through analyses from the perspectives of transformer and magnetic permeability. 

\section{Experimental setup}\label{exp}
\begin{figure}[ht]
\begin{center}
\includegraphics[width=0.75\textwidth,angle=0]{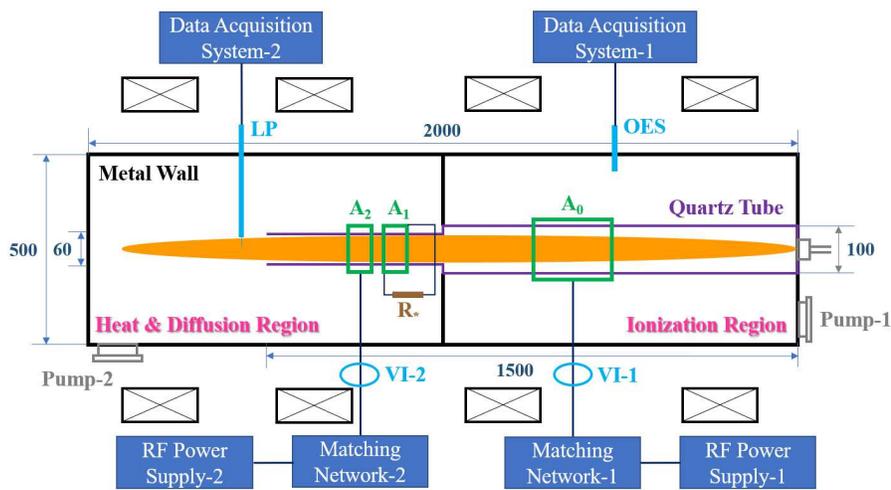}
\end{center}
\caption{Schematic of experimental setup: upgrade from PANTHER (Physics ANd Thruster oriented HElicon Research)\cite{Yang:2021aa} with second-stage antennas.}
\label{fg1}
\end{figure}

We conduct the experiments on PANTHER (Physics ANd Thruster oriented HElicon Research)\cite{Yang:2021aa}, which was built recently and produced blue-core helicon plasma, with an upgrade of second-stage antennas. Figure~\ref{fg1} shows a schematic. It has four main parts: vacuum chamber, power supplies, solenoid coils and diagnostics. The vacuum chamber comprises an outer stainless steel cylinder in length of $2$~m and diameter of $0.5$~m, which is equally and hermetically separated in the axial direction to form source region and heat \& diffusion region, respectively, and an inner quartz tube in length of $1.5$~m and diameter of $0.1$~m. The vacuum environment surrounding the antenna in the source region, which represents a special feature of PANTHER for space-related research, is maintained by rotary (AVT-$16.7$~L/s) and turbo-molecular (CBVAC-$1600$~L/s) pumps, while the low-pressure environment inside the quartz tube and the heat \& diffusion region is pumped by rotary (KYKY-$14$~L/s) and turbo-molecular (KYKY-$1600$~L/s) pumps. The pressure is monitored by vacuum gauges (Leybord PTR 90N). The working gas (helium throughout the paper) is fed into the quartz tube from the upstream of source region, and controlled via a series of mass-flow controllers in range of $0\sim 200$~sccm. The background pressure in level of $10^{-5}$~Pa can be achieved, whereas the working pressure is around $10^{-1}$~Pa. The power supplies consist of $13.56$~MHz-$1000$~W and $2\sim 30$~MHz-$500$~W. They are connected through matching networks to a half-turn helical antenna ($A_0$) for plasma generation in the source region and two isolated loop antennas ($A_1$ and $A_2$) with phase difference of $\pi/2$ for heating in the diffusion region, respectively. Please note that for the initial study of coupling physics in the present work, only one loop ($A_2$) is powered, while the other loop ($A_1$) is shortened by a resistance ($R_\ast=1~\Omega$) to measure the purely induced voltage and current; moreover, the driving frequency is fixed to $13.56$~MHz for both antennas, to simplify the wave activity and focus on the coupling process. There are four magnetic coils surrounding the vacuum chamber, which are powered and cooled separately to provide equilibrium magnetic field with adjustable profile. The magnetic field strength is up to $0.18$~T. The diagnostics include radio-frequency compensated Langmuir Probe (LP) to measure the electron density and temperature, two voltage-current (VI) probes to monitor the output from matching network to antenna, and Optical Emission Spectrometer (OES) to characterise the intensity and width of light emission which can then be used to infer the plasma parameters. 

\section{Data \& Analyses}\label{dat}
\subsection{Coupling without plasma}\label{vacuum}
\begin{figure}[ht]
\begin{center}
\includegraphics[width=0.5\textwidth,angle=0]{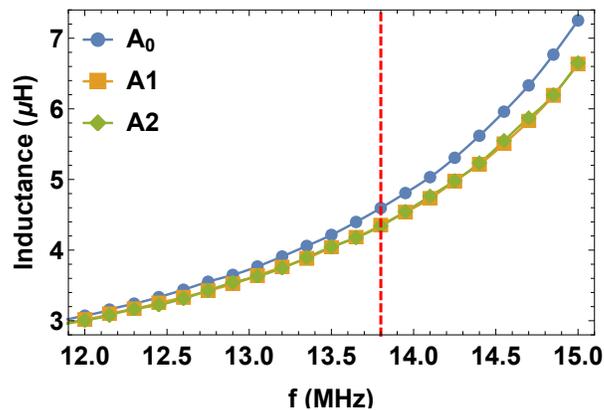}
\end{center}
\caption{Inductance of individual antennas for typical driving frequency range. Red dashed line labels the particular frequency of $13.56$~MHz employed in following sections. }
\label{fg2}
\end{figure}

\begin{figure}[ht]
\begin{center}
\includegraphics[width=0.5\textwidth,angle=0]{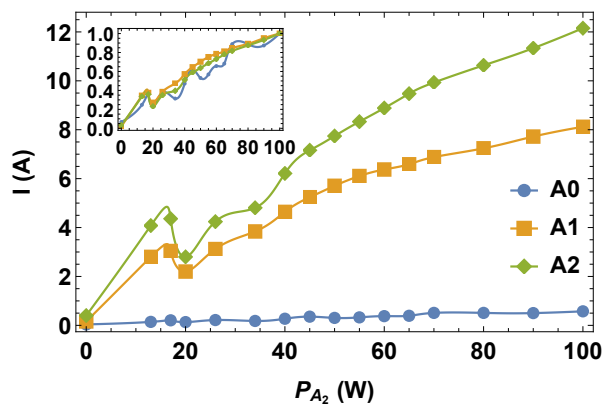}
\end{center}
\caption{Variation of antenna current on input power without plasma discharge. Inset gives normalised results, showing the same phase of current on $A_0$, $A_1$ and $A_2$.}
\label{fg3}
\end{figure}

\begin{figure}[ht]
\begin{center}
\includegraphics[width=0.5\textwidth,angle=0]{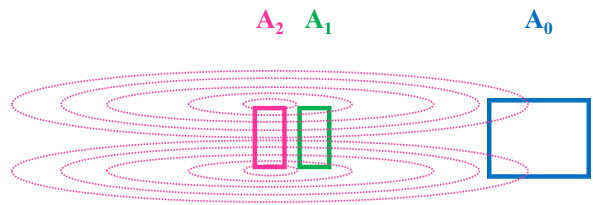}
\end{center}
\caption{Schematic of magnetic field lines generated by $A_2$.}
\label{fg4}
\end{figure}

We first measured the inductance of individual antennas for typical driving frequency range through a network analyser (KEYSIGHT-$4$~GHz), and carried it out in vacuum environment and without discharge. Figure~\ref{fg2} shows the results. It can be seen that the inductance grows monotonically with driving frequency, as expected, and the inductance of $A_0$ is slightly bigger than those of $A_1$ and $A_2$, especially in high frequency range, due to larger diameter, more turns, and longer axial extension. For the particular frequency of $13.56$~MHz employed in following sections, the inductance becomes $L_{A_0}=4.278$~$\mu$H, $L_{A_1}=4.083$~$\mu$H and $L_{A_2}=4.102$~$\mu$H, respectively. 
Then we switched on the power supply on $A_2$ while keeping the vacuum and no-discharge environment, and monitored the induced currents on neighbouring antennas. Figure~\ref{fg3} presents the variation of antenna current with input power exerted on $A_2$. We can see that the antenna currents all increase with input power, as expected, and the induced currents on $A_1$ and $A_0$ stay largely in the same phase of that on $A_2$, which is illustrated more clearly by the normalised results shown in the inset. The slight jump around $P_{A_2}=20$~W may be correlated to mode transition from isolated two-antenna interactions to integrated three-antenna interactions when the power exceeds a certain level. Indeed, the coupling between these antennas is eventually a three-body interaction problem, for example, the induced current on $A_0$ results from coupling with both $A_2$ and $A_1$, leading to interference feature displayed by the inset. Please note that the ratios between the antenna currents do not vary much and on average they are $I_{A_0}/I_{A_1}\approx 0.073$, $I_{A_0}/I_{A_2}\approx 0.047$ and $I_{A_1}/I_{A_2}\approx 0.699$, respectively.
To explore the underlying physics, we analyse the induced current from the perspective of transformer. Figure~\ref{fg4} shows a schematic of magnetic field lines generated by $A_2$. We can see that these antennas do not share the same magnetic flux, due to spacing and leakage. 
We assume that the magnetic flux enclosed by each antenna is $\Psi_0$ for $A_0$, $\Psi_1$ for $A_1$ and $\Psi_2$ for $A_2$. Since they are driven by the same power supply with fixed frequency of $f_\ast=13.56$~MHz, they all vary in form of $\exp(i \omega_\ast t)$ with $\omega_\ast=2\pi f_\ast$. According to the Faraday's law, the induced voltage $U$ is determined by time-varying magnetic field flux, i. e. 
\begin{equation}
U=\oint_l \mathbf{E}\cdot d\mathbf{l}=-\int_s \frac{\partial \mathbf{B}}{\partial t}\cdot d \mathbf{s}. 
\end{equation}
Here, $\mathbf{E}$ and $\mathbf{B}$ are time-varying electric and magnetic fields, respectively, and $l$ and $s$ are the path and area of integrations. Referring to transformer, we write the equations: 
\begin{equation}
U_0=R_\ast I_0=-n_0\frac{d \Psi_0}{d t}=-i\omega_\ast n_0 \Psi_0,
\end{equation}
\begin{equation}
U_1=R_\ast I_1=-n_1\frac{d \Psi_1}{d t}=-i\omega_\ast n_1 \Psi_1,
\end{equation}
\begin{equation}
U_2=R_\ast I_2=-n_2\frac{d \Psi_2}{d t}=-i\omega_\ast n_2 \Psi_2,
\end{equation}
with $n$ the number of coils for each antenna. Therefore, the ratios of antenna current equals to the ratios of induced magnetic flux ($n\Psi$), namely
\begin{equation}
I_0/I_1/I_2=n_0\Psi_0/n_1\Psi_1/n_2\Psi_2.
\end{equation}
For the loop antennas of $A_1$ and $A_2$, we have $n_1=n_2= 1$; while for the half-turn helical antenna of $A_0$, we may write $n_0\approx 2.5$. It can be seen that although $A_1$ is very close to $A_2$, it only captures $69.9\%$ of the magnetic flux generated by $A_2$; while for $A_0$, the percentage is much lower ($4.7\%/2.5$) because of much longer distance. 

\subsection{Coupling with plasma}\label{discharge}
\begin{figure}[ht]
\begin{center}
\includegraphics[width=0.5\textwidth,angle=0]{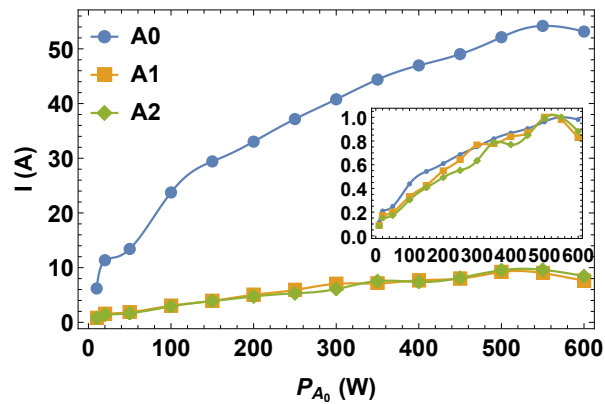}
\end{center}
\caption{Variation of antenna current on input power with plasma discharge. Inset gives normalised results, showing the same phase of current on $A_0$, $A_1$ and $A_2$.}
\label{fg5}
\end{figure}

\begin{figure}[ht]
\begin{center}
\includegraphics[width=0.9\textwidth,angle=0]{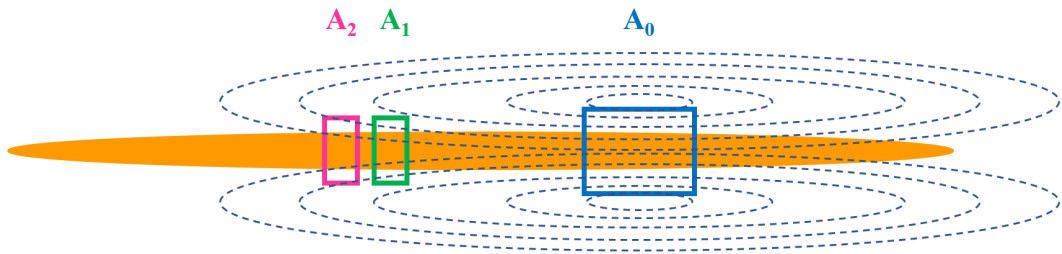}
\end{center}
\caption{Schematic of magnetic field lines generated by $A_0$.}
\label{fg6}
\end{figure}

We show above that the coupling between antennas immersed in vacuum environment can be described simply by the Faraday's equation. This situation, however, becomes complicated when there is a plasma cylinder going through them, as shown in Fig.~\ref{fg1}. To reveal the underlying physics, we pumped in helium with background pressure of $P_B=0.042$~Pa, and applied external magnetic field of strength $B_0=0.08$~T. Figure~\ref{fg5} shows the variation of antenna current with input power exerted on $A_0$. Please note that for all the cases of $P_{A_2}=0$~W throughout the paper, the antenna $A_2$ is shortened by a resistance ($R_\ast=1~\Omega$). This also applies to $A_1$ for which $P_{A_1}=0$~W applies. 
We find that the current increases with input power for all antennas under the same phase, consistent with Fig.~\ref{fg3}. The ratios between them are independent of input power and they are $I_{A_1}/I_{A_0}\approx 0.150$, $I_{A_2}/I_{A_0}\approx 0.146$ and $I_{A_2}/I_{A_1}\approx 0.973$, respectively. Compared with Fig.~\ref{fg3}, one can see that the ratios are much higher, indicating better coupling between antennas with the existence of plasma. Since the induced power origins from $A_0$ and the distance between $A_1$ and $A_0$ is shorter than that between $A_2$ and $A_0$, it is reasonable to have $I_{A_2}/I_{A_1}<1$. Again, we draw the magnetic flux surrounding these antennas (Fig.~\ref{fg6}), and analyse these ratios referring to transformer but with the inclusion of plasma. As pointed in early references\cite{Grad:1971aa, Bodin:1976aa}, a stabilised plasma can be paramagnetic, and could increase the magnetic permeability and confine the magnetic flux, behaving similarly to the magnetic core for a transformer. This surely enhances the coupling effect among antennas and thereby the ratios of induced current. Please note that although plasma cannot be treated as normal magnetic material\cite{Chen:2016aa}, the gyrating motions of charged particles do provide additional magnetic moment, in form of $\mu_m=m v_\perp^2/2B$, which leads to the increased effective magnetic permeability ($\mu_{e}=\mu_0+\mu_m$). 

\begin{figure}[ht]
\begin{center}$
\begin{array}{ll}
(a)&(b)\\
\includegraphics[width=0.45\textwidth,angle=0]{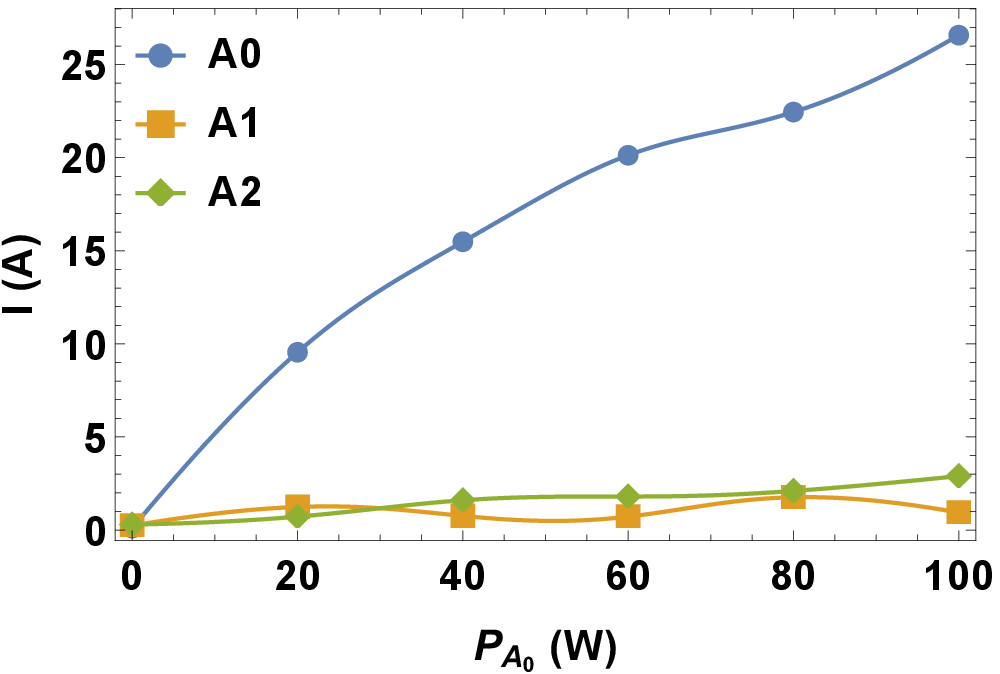}&\includegraphics[width=0.45\textwidth,angle=0]{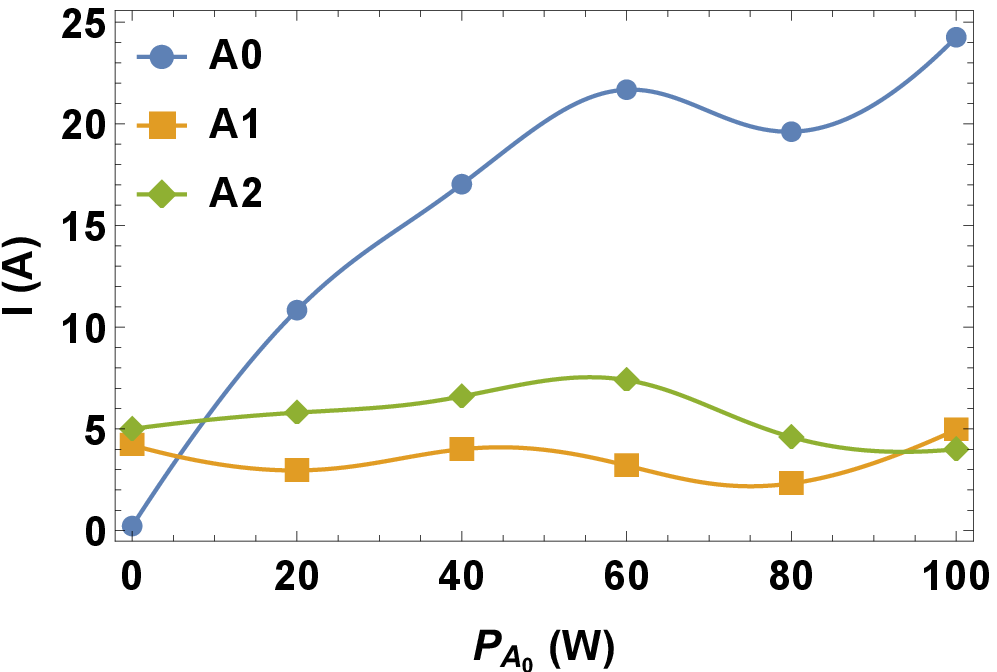}
\end{array}$
\end{center}
\caption{Variation of antenna current with input power without external magnetic field ($B_0=0$~T): (a) $P_{a_2}=0$~W, (b) $P_{a_2}=20$~W.}
\label{fg7}
\end{figure}

\begin{figure}[ht]
\begin{center}$
\begin{array}{ll}
(a)&(b)\\
\includegraphics[width=0.45\textwidth,angle=0]{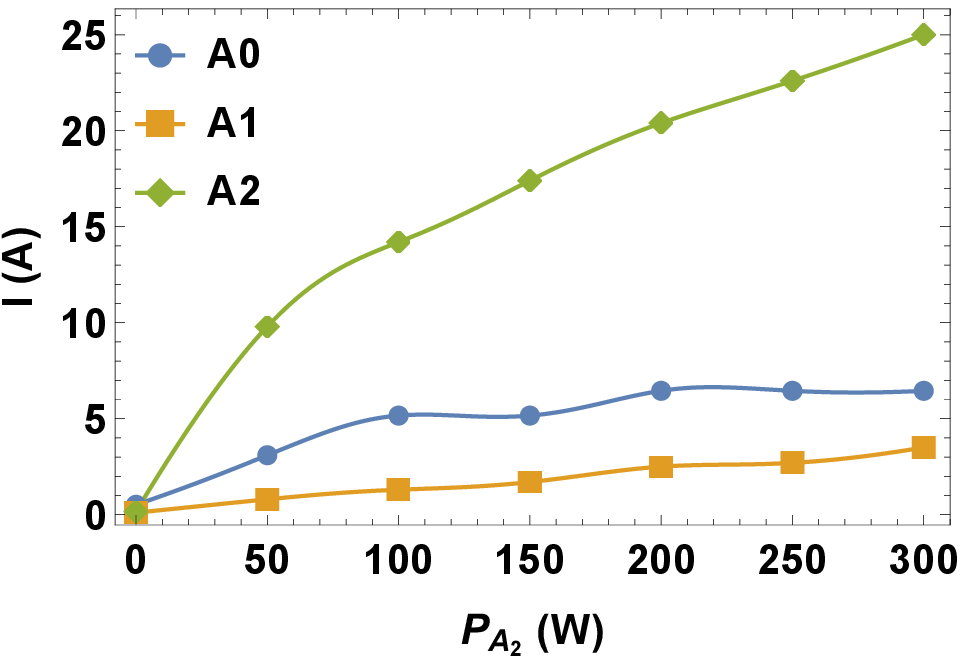}&\includegraphics[width=0.45\textwidth,angle=0]{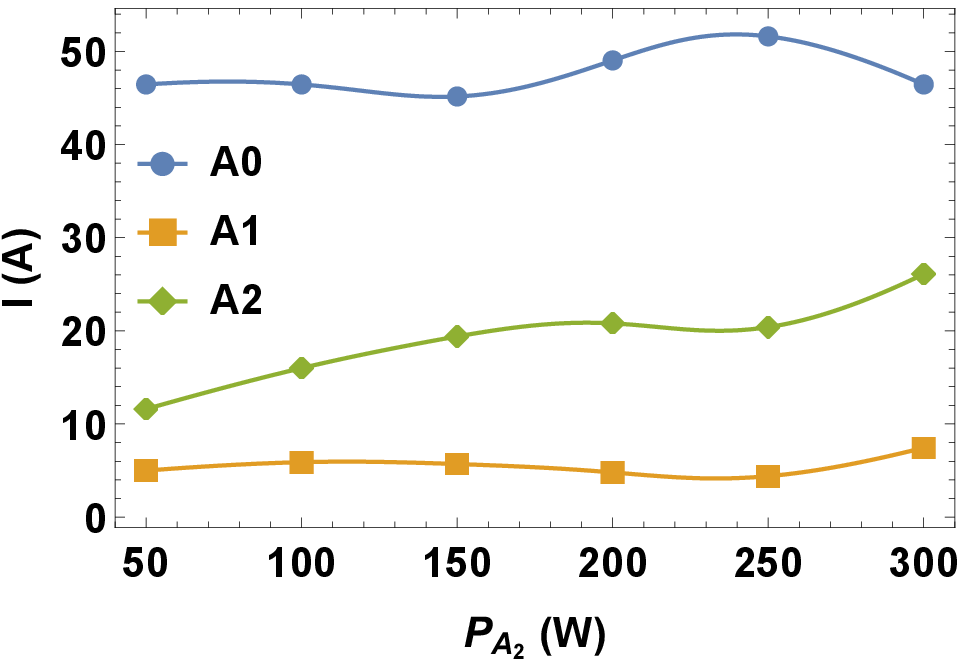}
\end{array}$
\end{center}
\caption{Variation of antenna current with input power with external magnetic field ($B_0=0.12$~T): (a) $P_{A_0}=0$~W, (b) $P_{A_0}=500$~W.}
\label{fg8}
\end{figure}

\begin{figure}[ht]
\begin{center}
\includegraphics[width=0.9\textwidth,angle=0]{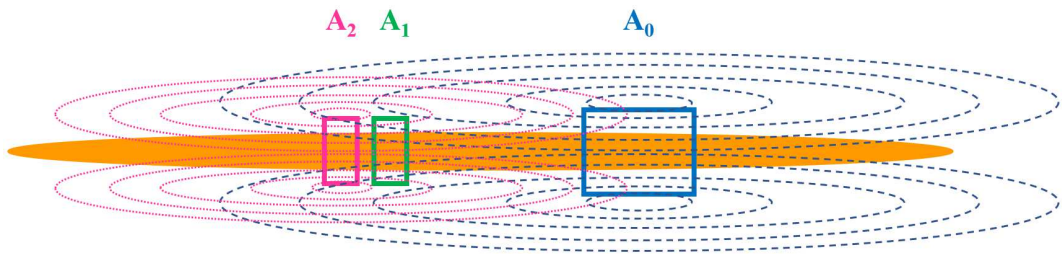}
\end{center}
\caption{Schematic of magnetic field lines generated by $A_2$ and $A_0$.}
\label{fg9}
\end{figure}

\begin{figure}[ht]
\begin{center}$
\begin{array}{ll}
(a)&(b)\\
\includegraphics[width=0.45\textwidth,angle=0]{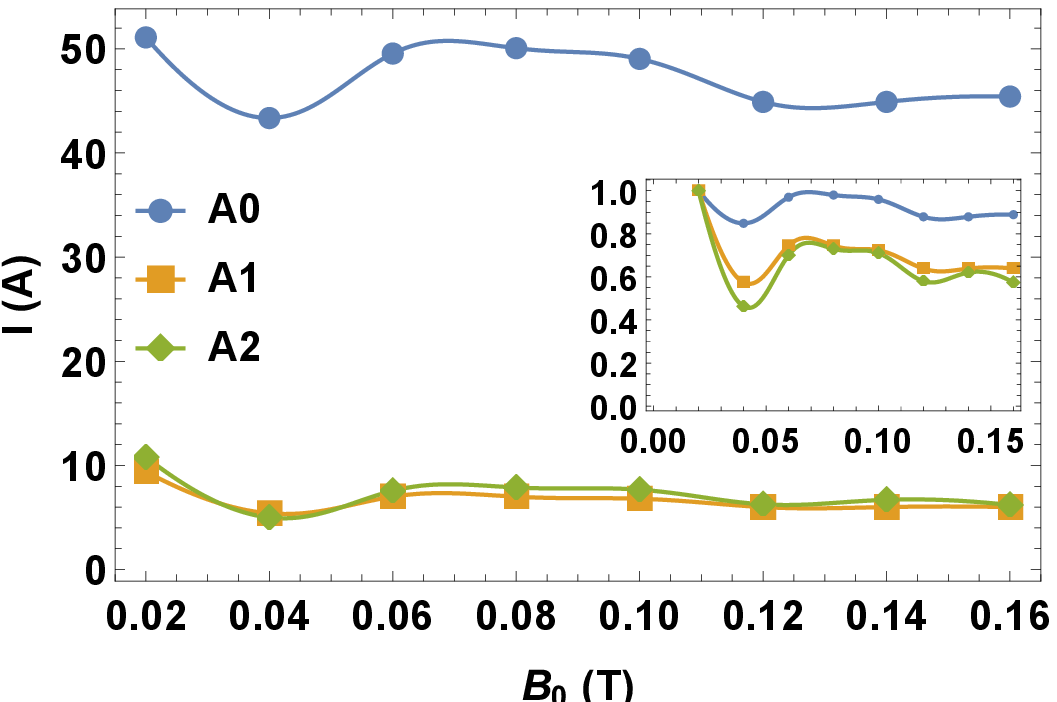}&\includegraphics[width=0.45\textwidth,angle=0]{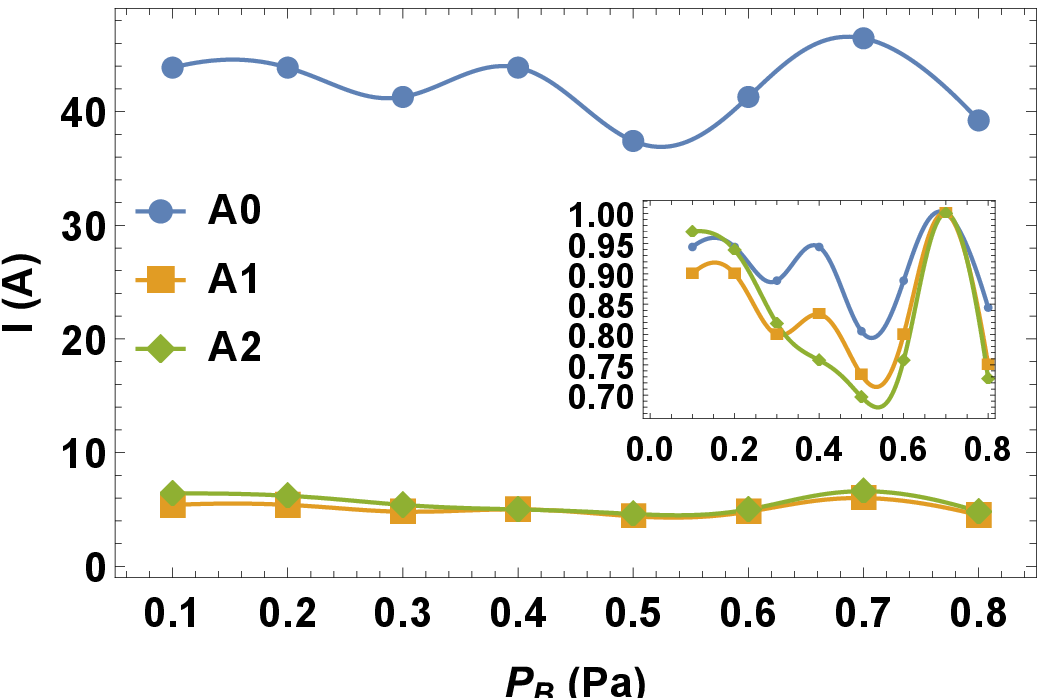}
\end{array}$
\end{center}
\caption{Variation of antenna current with: (a) external magnetic field ($P_B=0.02$~Pa), and (b) background pressure ($B_0=0.08$~T) for $P_{A_0}=500$~W and $P_{A_2}=0$~W.}
\label{fg10}
\end{figure}

Then we switched on the power supply for two antennas, i. e. $A_0$ and $A_2$, and compared with one-antenna discharge. Experiments were conducted for unmagnetised and magnetised cases, respectively. Figure~\ref{fg7} shows the variation of antenna current with input power on $A_0$ without external magnetic field ($B_0=0$~T). The averaged ratios are: (a) $I_{A_0}/I_{A_1}\approx 16.249$, $I_{A_0}/I_{A_2}\approx 9.055$ and $I_{A_1}/I_{A_2}\approx 0.763$; (b) $I_{A_0}/I_{A_1}\approx 4.682$, $I_{A_0}/I_{A_2}\approx 2.959$ and $I_{A_1}/I_{A_2}\approx 0.689$. It can be seen that the addition of input power on $A_2$ slightly lowers the current on $A_0$, which is consistent with the diamagnetic induction between these two antennas. While the current on $A_2$ increases as expected, the current on $A_1$ also increases. This is because the coupling between $A_1$ and $A_2$ is stronger than that between $A_1$ and $A_0$, due to much shorter distance. Please note that for the case of $P_{a_2}=0$~W, the ratio of $I_{A_1}/I_{A_2}$ is ideally bigger than unity if simply considering the distance; however, the mutual induction transits from two-body to three-body problems when the power becomes higher, making the induction more complicated. This also applies to the case of $P_{A_0}=0$~W shown in Fig.~\ref{fg8}, namely the ratio of $I_{A_1}/I_{A_0}$ which also involves the number of antenna coils ($n_0/n_1\approx 2.5$).
For the magnetised case of $B_0=0.12$~T, the variation of antenna current with input power on $A_2$ is given by Fig.~\ref{fg8}. The averaged ratios are: (a) $I_{A_0}/I_{A_1}\approx 3.236$, $I_{A_0}/I_{A_2}\approx 0.723$ and $I_{A_1}/I_{A_2}\approx 0.186$; (b) $I_{A_0}/I_{A_1}\approx 8.878$, $I_{A_0}/I_{A_2}\approx 2.650$ and $I_{A_1}/I_{A_2}\approx 0.304$. Similar to Fig.~\ref{fg7}, the addition of input power on $A_0$ not only increases the current magnitude of $A_0$, as expected, but also increases the induced current on the unpowered $A_1$. However, the addition of external magnetic field does not show clear trend, which will be confirmed by the parameter studies in Fig.~\ref{fg10}(a). To understand the effects of additional power, we attribute them to the formed plasma of higher density and temperature, which increase the number and energy of charged particles, respectively. This yields larger effective magnetic moment ($\sum{\mu_m=\sum{m v_{\perp}^2/2 B}}$) and thereby stronger coupling. 
The absolute magnitude of antenna current is eventually determined by the three-body inductions, as shown in Fig.~\ref{fg9}; the induced current on $A_1$ is a result from both $A_0$ and $A_2$, although the mutual induction between $A_0$ and $A_2$ is less effected by the unpowered $A_1$. 

\begin{figure}[ht]
\begin{center}
\includegraphics[width=0.9\textwidth,angle=0]{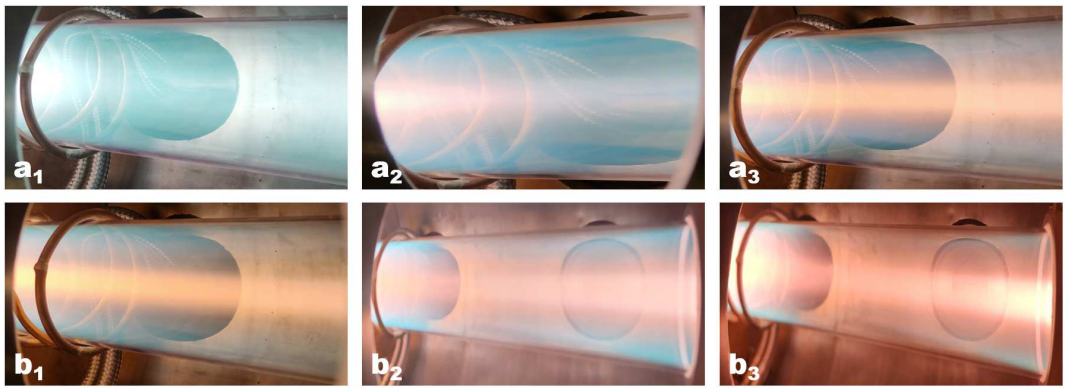}
\end{center}
\caption{Discharge images for varied external magnetic field ($a_1$: $0.02$~T, $a_2$: $0.04$~T, $a_3$: $0.14$~T) and background pressure ($b_1$: $0.1$~Pa, $b_2$: $0.5$~Pa, $b_3$: $0.8$~Pa).}
\label{fg11}
\end{figure}

Finally, we studied the effects of magnetic field and background pressure on antenna coupling. The applied input power is $P_{A_0}=500$~W and $P_{A_2}=0$~W. Figure~\ref{fg10} shows the measurements of antenna current. The averaged ratios are: (a) $I_{A_0}/I_{A_1}\approx 7.180$, $I_{A_0}/I_{A_2}\approx 6.738$ and $I_{A_1}/I_{A_2}\approx 0.934$; (b) $I_{A_0}/I_{A_1}\approx 8.399$, $I_{A_0}/I_{A_2}\approx 7.744$ and $I_{A_1}/I_{A_2}\approx 0.921$. We can see that, especially from the normalised results, the current magnitude drops first and then increases to a stable value when the field strength is increased, minimising around $B_0=0.04$~T; while as the pressure level is enhanced, the current magnitude drops gradually with oscillations, peaking around $P_B=0.7$~Pa for the pressure range employed. Assuming the microcosmic electron spin that determines the magnetic permeability remains unchanged, the macroscopic effects of field and pressure take place only through the magnetic moment of plasma, i. e. $\mu_m=m v_\perp^2/2B$, either directly (field) or indirectly (pressure) through collisions. 
Figure~\ref{fg11} displays the discharge images for typical field strengths and pressure levels. One can see that the plasma is more confined near axis for higher field and lower pressure. Detailed calculation of the paramagnetic or diamagnetic property of cylindrical plasma requires three-dimensional profiles of plasma density and temperature, which are missing at the moment. We thereby leave it as a separate study and will present in the future. 

\section{Conclusion}\label{con}
To reveal the mutual induction between co-axial antennas wrapping around a plasma cylinder, measurements are carried out for their induced currents for both unmagnetised and magnetised cases. We find that the induced currents evolute in the same phase, and their magnitudes are determined eventually by the multi-antenna inductions through Faraday's law. The separation distance between antennas and the existence of plasma column effect these inductions remarkably, via the leakage of magnetic flux and the enhanced magnetic permeability respectively. The additional effect of external magnetic field behaves through the ionisation process, plume confinement and magnetic moment, whereas the background pressure effects the ionisation and magnetic moment largely through collisions. Analyses referring to transformer provide intuitive understanding of this coupled system. In this respect, the plasma column acts as a magnetic core for a transformer, enhancing the coupling effect by limiting the path of magnetic flux. The magnetic property of plasma column, however, relies on three-dimensional profiles of density and temperature, and is left for detailed calculation in the future. 

\ack
This work is supported by the Chinese Academy of Sciences “$100$ Talent” Program (B), the Science Foundation of Institute of Plasma Physics (No. DSJJ-$2020$-$07$), and the Shanghai Engineering Research Center of Space Engine (No. $17$DZ$2280800$). 

\section*{References}
\bibliographystyle{unsrt}

\end{document}